\documentclass[twocolumn,aps,pre,showpacs,amsmath,amsfonts,amssymb,floatfix]{revtex4}
\usepackage{morefloats}
\usepackage{amsmath}
\usepackage{graphicx}
\usepackage{dcolumn}
\usepackage{bm}
\usepackage{color}
\definecolor{orange}{rgb}{1,0.5,0}
\begin{document}
\title{Role of delay in the mechanism of cluster formation}

\author{Aradhana Singh}
\affiliation{ Complex Systems Lab, Indian Institute of Technology Indore,
IET-DAVV Campus Khandwa Road, Indore-452017}
\author{Sarika Jalan\footnote{sarika@iiti.ac.in} }
\affiliation{ Complex Systems Lab, Indian Institute of Technology Indore,
IET-DAVV Campus Khandwa Road, Indore-452017}
\author{J\"urgen Kurths}
\affiliation{Potsdam Institute for Climate Impact Research, P.O. Box 601203, D-14412 Potsdam}
\affiliation{Institute for Complex Systems and Mathematical Biology, University of Aberdeen,
Aberdeen-AB243FX}
\begin{abstract}
We study the role of delay in phase synchronization and 
phenomena responsible for cluster formation in delayed 
coupled maps on various networks. Using numerical simulations, we 
demonstrate that the presence of delay may  
change the mechanism of unit to unit interaction.
At weak coupling values, 
same parity
delays are associated with the same phenomenon of cluster formation and exhibit similar dynamical evolution.
Intermediate coupling values yield rich delay-induced driven cluster patterns.
A Lyapunov function analysis sheds
light on the robustness of the driven clusters observed for delayed bipartite 
networks. Our results reveal that delay may lead to a completely different 
relation, between dynamical 
and structural clusters, than observed for the undelayed case.
\end{abstract}
\pacs{05.45.Xt,05.45.Pq}
\maketitle
Studying the impact of network topology on dynamical processes is of fundamental importance for 
understanding the functioning of many real world complex networks \cite{rev-network}. The dynamical behavior of a system 
depends on the collective behavior of its individual units. 
One of the most fascinating emergent behavior of 
interacting chaotic units is the observation of synchronization \cite{book_Kurths}. 
In general, synchronization may lead to more complicated
patterns including clusters \cite{SJ_prl2003,pattern_Kestler,cluster_laser_brain}. 
The interplay between underlying network structure and dynamical clusters has been the prime area of focus 
for the last two decades \cite{functional_net}.  
Furthermore, communication delay naturally arises in extended systems \cite{book_delay}. A delay gives rise to 
many new phenomena in dynamical systems such
as oscillation death, stabilizing periodic orbits,
enhancement or suppression of synchronization, chimera state, etc
\cite{osc_death_delay,Atay_analytical,Thilo,delay_enhance_syn,delay_coup_osc,delay_suppress_syn,chimera}. 

In this paper, we study the impact of delay on the phenomenon of
phase synchronized clusters in coupled map networks. 
We investigate the formation of clusters
on various networks namely, 1-d lattice, small-world, random, scale-free and 
bipartite networks \cite{network_algorithm}, and provide
a Lyapunov function analysis for bipartite networks to explain possible 
reasons behind the role of a delay on synchronized clusters.
So far, studies on delayed 
coupled dynamical systems
mostly concentrated on a global synchronized state, except a few recent studies which have focused on  pattern formation or clustered states
\cite{pattern_Kestler,cluster_laser_brain,delay_msf,cluster_Lakshmanan}.
These studies have revealed that delay emulates qualitative changes in clustered state, whereas
mechanism of delayed unit to unit interactions 
has not been investigated so far.

Previous studies on undelayed coupled systems have identified two different
phenomena for synchronization namely, the driven (D) and the self-organized 
(SO) \cite{SJ_prl2003}. SO (D) synchronization refer to the state
when clusters are formed because of intra-cluster (inter-cluster) couplings.
Here, we report that a delay can play a crucial role in the formation 
of clusters as well as the phenomenon behind it.  
The formation of delay-induced synchronized clusters may be 
because of inter-cluster couplings, instead of coupling between synchronized
units \cite{cluster_laser_brain,delay_msf}. 
Introduction of a delay may result in a transition from D to  
SO synchronization or vice versa.
Furthermore,  our studies demonstrate a delay-induced emergence of dynamical
phase synchronized D patterns. These patterns are stable with time and 
are dynamical with respect to a change in $\tau$.
A delayed bipartite network leads to a transition from SO to D 
synchronization in an intermediate coupling range irrespective of $\tau$.

Here we take networks with a less average degree ($N_C \sim N$), leading to 
phase synchronized clusters instead of a complete synchronized state which usually spans all the nodes.
We consider a network 
of $N$ nodes and $N_c$ connections between the nodes. Let each node of the network
be assigned a dynamical variable $x^i, i = 1, 2, \hdots, N$. 
The dynamical evolution is defined by the well 
known coupled maps \cite{rev_cml},
\begin{equation}
x_i(t+1) = (1-\varepsilon) f(x_i(t)) + \frac{\varepsilon}{k_i} \sum_{j=1}^N A_{ij} g(x_j(t - \tau))
\label{cml}
\end{equation}
Here $A$ is the adjacency matrix with elements
$A_{ij}$ taking values $1$ and $0$ depending upon whether there is a
 connection between $i$ and $j$ or not. $ k_{i}$ = $\sum_{j=1}^{N}A_{ij}$ is the degree
of the $ith$ node and $\varepsilon$ is the overall coupling constant.
In the present investigation we consider a homogeneous delay $\tau$.
The function $f(x)$ defines a local nonlinear map, $g(x)$
defines the nature
of coupling between the nodes. 
We consider phase synchronization as described in \cite{Phase_syn}. 
\begin{figure}[t]
\begin{center}
\includegraphics[width=0.32\columnwidth,height=3.8cm]{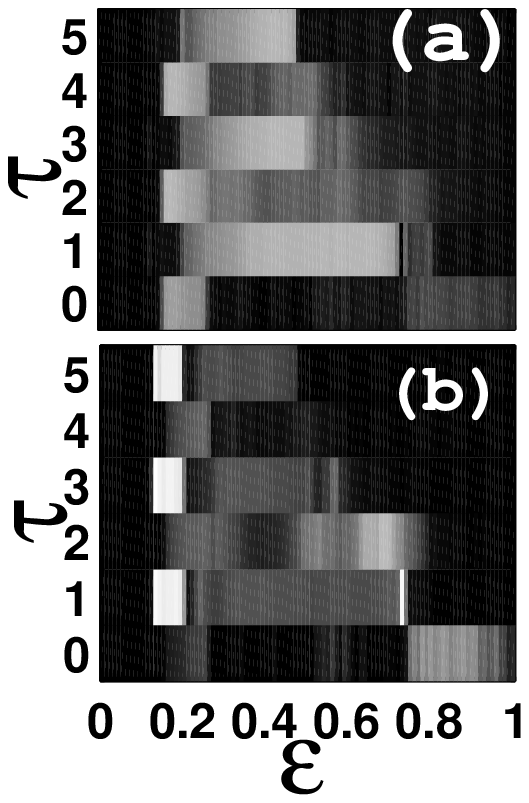}
\includegraphics[width=0.28\columnwidth,height=3.8cm]{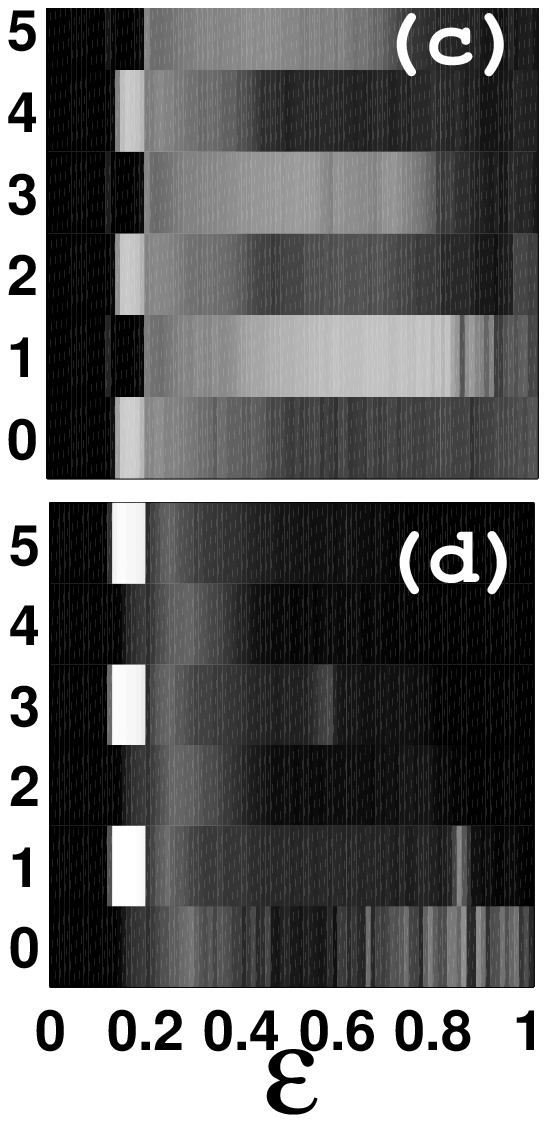}
\includegraphics[width=0.32\columnwidth,height=3.8cm]{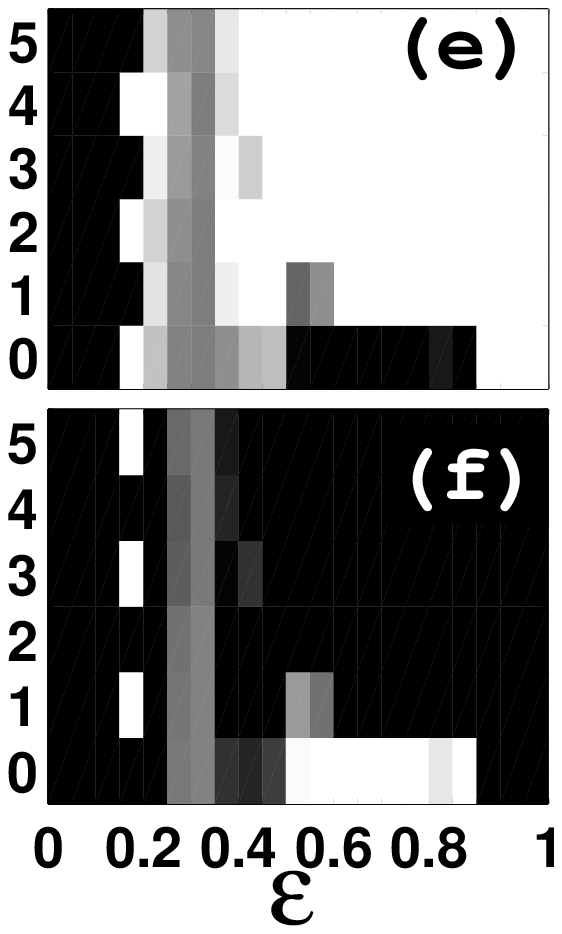}
\end{center}
\caption{Phase diagram of phase synchronization patterns in system (\ref{cml}) for 1-d lattice with $N=50$, $<k>=4$.  Gray-scale
encoding represents  values of (a) $f_{inter}$ and (b) $f_{intra}$
Local
dynamics is governed by logistic map $f(x)=4x(1-x)$ and coupling function
$g(x)=f(x)$.
The figure is obtained by averaging over 20 random initial conditions. 
The regions, which are black in both graphs
(a) and (b), correspond to states of no cluster formation.
Both subfigures with gray shades correspond to
clusters having both inter- and intra-couplings. The regions in (a), which are lighter as compared to
the corresponding $\varepsilon$ and $\tau$ values in (b),
refer to dominant D phase synchronized clusters,
and the reverse refer to dominant SO phase synchronized clusters. White regions
in (a) and (b) refer to ideal D or ideal SO clusters  respectively.
The regions, which are dark gray in (a) and black in (b) or vice-versa, correspond to states where
a much less clusters are formed.
(c) and (d) are for scale-free, and (e) and (f) are for bipartite networks and 
demonstrate the same as (a) and (b) respectively.}    
\label{Fig_Phase_NNC_finter_fintra_Delay}
\end{figure}
As the network evolves, it splits into several synchronized clusters. 
In order to have a clear picture of SO and D behavior,
we use $f_{intra}$ and $f_{inter}$ measures for intra- and inter-cluster couplings as follows 
\cite{SJ_prl2003};
$f_{intra} = N_{intra}/N_c$ and  $f_{inter} = N_{inter}/N_c$.
where $N_c$ is total number of connections in the network, 
$N_{intra}$ and $N_{inter}$ are the numbers of intra- and inter-cluster couplings, 
respectively \cite{note_inter_intra}.
We evolve Eq.(\ref{cml}) starting from random initial conditions, and study the dynamical
behavior of nodes after an initial transient.
First let us consider local dynamics being governed by the logistic map $f(x) = 4x(1-x)$,
and the coupling function $g(x)=f(x)$.\\
The undelayed coupled maps on all model networks we have considered 
yield  dominant D clusters in
the range ($0.16 \lesssim \varepsilon \lesssim 0.25$).
For rest $\varepsilon$ values, coupled maps on 1-d lattice and 
small-world networks exhibit no phase synchronization, except for $\varepsilon \gtrsim 0.74 $ having
mixed clusters with very small values of $f_{inter}$ and $f_{intra}$ 
(Figs.~\ref{Fig_Phase_NNC_finter_fintra_Delay}(a) and \ref{Fig_Phase_NNC_finter_fintra_Delay}(b)).
In this $\varepsilon$ range scale-free and random networks favor synchronization 
yielding better cluster
formation than the corresponding regular and small-world networks 
(Figs.~\ref{Fig_Phase_NNC_finter_fintra_Delay}(c) and \ref{Fig_Phase_NNC_finter_fintra_Delay}(d)), 
while bipartite networks lead to ideal SO synchronization 
for $0.45 \lesssim \varepsilon \lesssim 0.85$ and 
ideal D synchronization for higher $\varepsilon$ values \cite{SJ_prl2003}.
Upon introducing a delay of $\tau=1$ in Eq.(\ref{cml}),
after very small $\varepsilon$ values, 
for which there is no phase synchronization for the undelayed case
(black color for the Figs.~\ref{Fig_Phase_NNC_finter_fintra_Delay}(a), (b), (c) 
and (d)), we get SO clusters in the 
region $0.13 \lesssim \varepsilon \lesssim 0.2$ 
as seen from the white regions in the Figs.~\ref{Fig_Phase_NNC_finter_fintra_Delay}(b) 
and \ref{Fig_Phase_NNC_finter_fintra_Delay}(d). For most of the $\varepsilon$ values 
in this region, coupled dynamics exhibits a periodic evolution
with a period depending upon $\tau$.
For a further increase in $\varepsilon$, in the middle coupling range, 
the 1-d lattice, small-world, scale-free and random networks lead to an increase in D
synchronization in $0.4 \lesssim \varepsilon \lesssim 0.7$, whereas for complete bipartite networks, 
ideal D synchronization is achieved 
for almost all $\varepsilon$ values in this range.
For $0.85 \lesssim \varepsilon \lesssim 1.0$,
the delayed case exhibits a very small (almost negligible) cluster formation compared to 
 the undelayed case, hence indicating a suppression of synchronization for all the networks
except for bipartite networks forming ideal D clusters.
For $\tau=2$,  lower $\varepsilon$ range coerces the formation of dominant D clusters, similar to the 
undelayed case. As $\varepsilon$ increases, 1-d lattice and small-world networks lead to mixed clusters,
whereas scale-free and random networks lead to dominant D clusters.
Bipartite networks emulates ideal D synchronization. 
With a further increase in $\tau$, at a lower $\varepsilon$ range, 
odd $\tau$ leads to a similar 
behavior as for $\tau=1$ and even $\tau$ exhibits similar behavior as for $\tau=0$
and $\tau=2$.
For the intermediate $\varepsilon$ range there is a 
suppression in synchronization. Higher $\varepsilon$ 
values manifest no cluster formation as illustrated by the 
black regions in Fig.~\ref{Fig_Phase_NNC_finter_fintra_Delay} for all networks except bipartite which form ideal D clusters for $\varepsilon \gtrsim 0.4$ for all $\tau$.

Above description boils down to the following;
there is a $\varepsilon$ region which demonstrates a change in the phenomenon of cluster formation with a change in
$\tau$. 
The zero and even delays imply dominant D clusters,
whereas odd delays imply ideal or dominant SO clusters.
Moreover, odd delays lead to SO clusters with a periodic evolution, whereas
zero and even delays lead to D cluster with periodic, quasi-periodic
or the chaotic evolution  \cite{supp_mat}. Note that the measure of 
phase-synchronization considered here satisfies the metric properties, but
does not include anti-phase synchronization and consequently nodes
being anti-phase synchronized would land up in different clusters. 
However, anti-phase
or phase shift synchronization is not the only cause behind the separation of nodes in clusters
 \cite{supp_mat}.

Though the nodes in various clusters display a
rich dynamical evolution, a simple  
analysis for periodic synchronized state, for example bipartite networks in the lower 
$\varepsilon$ region, provides a basic understanding of different behaviors indicated by odd and even
delays.
 In this $\varepsilon$ region, the coupling term having a delay part yields:
 \[f(x(t-\tau)) = \left\{ 
\begin{array}{l l}
  f(p_{1}) & \quad \mbox{if $\tau=0$ and even }\\
  f(p_{2}) & \quad \mbox{if $\tau$ is odd, }\\ \end{array} \right. \]
implying that the discrete time delay considered here introduces a
difference on the evolution of the nodes (Eq.(\ref{cml})) depending upon the parity of delay,
 and
thus leading to a particular behavior for zero and even delays but
a different behavior for odd delays. 

Furthermore, a change in $\tau$ leads to a change in SO or D cluster pattern.  
A pattern refers to a particular phase synchronized state, containing 
information of all the pairs of the phase synchronized nodes distributed in the various clusters.
A change in the pattern refers to the state when members of a cluster get changed as an effect of delay.
For some cases we observe ideal D or SO clusters.
Ideal SO synchronization refers to a state when clusters do not have
any connection outside the cluster, except one. The ideal D synchronization refers to the state
when clusters do not have any connections within them, and all connections are
outside.

\begin{figure}[t]
\centerline{\includegraphics[width=0.7\columnwidth]{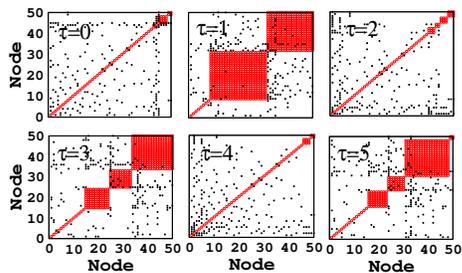}}
\caption{(Color online) A typical behavior of coupled dynamics illustrating
D patterns observed with changing $\tau$. 
Squares represent clusters, diagonal dots represent freely-evolved nodes 
while off-diagonal dots imply that the two corresponding nodes are coupled (i.e. $A_{ij}=1$).
In each case the node numbers are reorganized so that the nodes belonging to the same
cluster are numbered consecutively. 
The example presents a scale-free network 
with $N=50$ and $\varepsilon=0.6$.
For $\tau=0$, very few nodes are forming cluster.
For $\tau=1, 3$ and $5$, nodes form dominant
D clusters, whereas $\tau=2$ and $4$, yield very few nodes forming clusters of ideal D type.}
\label{Fig_SF_clus_0.6}
\end{figure}

Next we focus to the $\varepsilon$ range where
the delayed evolution leads to ideal D clusters for bipartite networks, and
dominant D clusters for other networks.
In bipartite networks, 
a division of nodes into ideal D clusters is unique. Whereas for other network structures, 
there can be various possible ways in which one can distribute nodes to form ideal D 
(for average degree two) or dominant D (for larger average degree) clusters. 
Fig.~\ref{Fig_SF_clus_0.6} plots snapshots of clusters for different $\tau$  
by keeping all other parameters same. 
It indicates that with a change 
in $\tau$, both nodes forming clusters as well as size of clusters are changed.
Note that the 
dynamical evolution here may be periodic, quasi-periodic or chaotic.
In this region, for a particular delay value, the clusters are almost 
stable with time evolution, with few nodes of the floating type \cite{SJ_prl2003}. But a 
change of $\tau$ has a drastic impact on cluster patterns,
and may lead to entirely different sets of nodes forming clusters.
Hence D patterns obtained in this range are dynamic with respect to a 
change in $\tau$. However, the phenomenon behind the pattern 
formation does not change, and 
the D mechanism is mainly responsible for the cluster formation.
For this $\varepsilon$ range, a delayed evolution on a bipartite network yields ideal D 
clusters for all $\tau$ values we have investigated.

\begin{figure}[t]
\includegraphics[width=0.64\columnwidth,height=1.8cm]{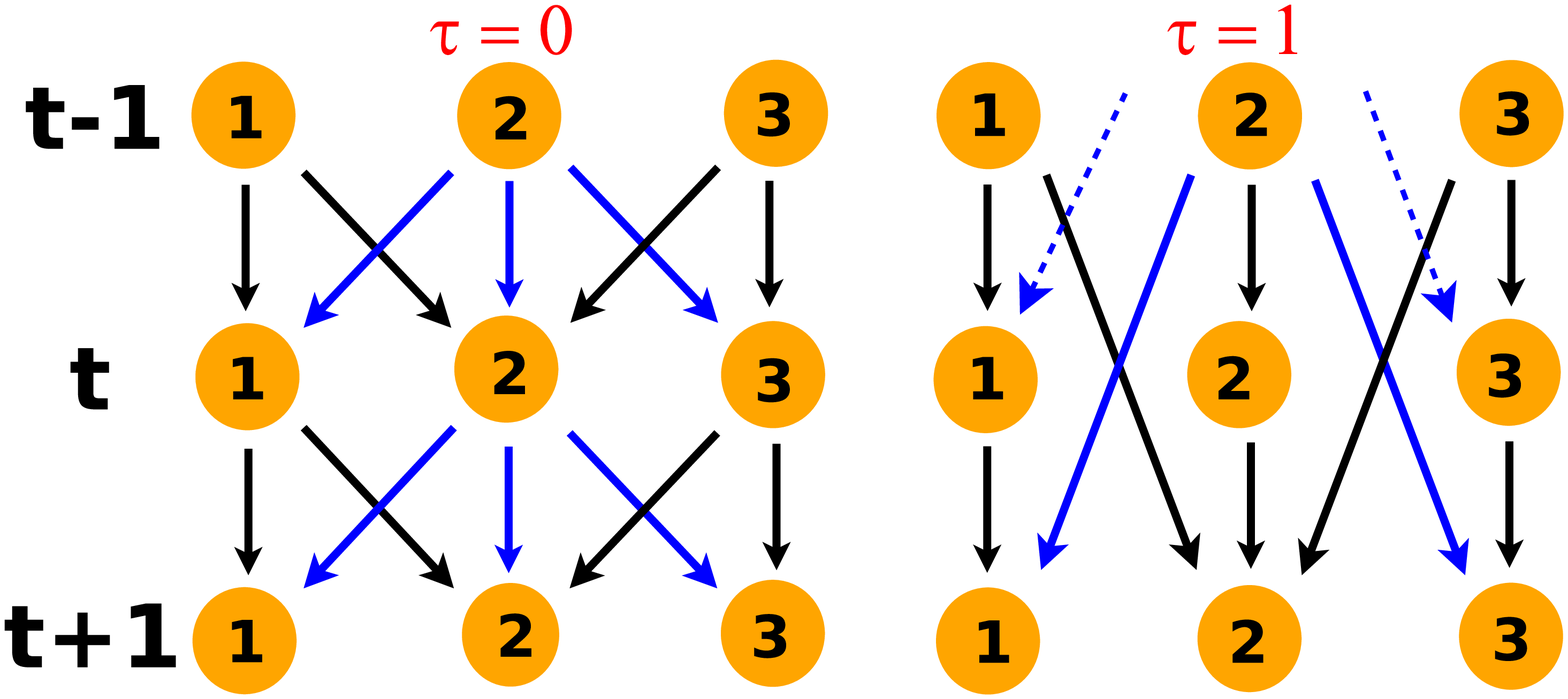}
\caption{(Color online) Three nodes 
schematic diagram illustrating impact of delay. Arrows 
depict the direction of information flow as governed by Eq.(\ref{cml}). 
The dashed lines show the flow of information from the $(t-2)^{th}$ time step.
For $\tau=0$, evolution of all nodes ({\color{orange}{$\bullet$}})
 receive information from the second node
(left panel), whereas in presence of delay, evolution of connected nodes at a particular
time do not involve any common term (right panel).  
For both panels, first and third nodes 
are connected with the second one leading to the construction of the smallest possible bipartite 
network.}
\label{fig_net_clus}
\end{figure}

Aforementioned can be explained further using the example of
bipartite networks. 
A Lyapunov function analysis can be carried out for 
delayed case in a very similar fashion as for $\tau=0$ described in \cite{physicaA2012},
and for a pair of synchronized nodes on a bipartite network can be written as:
\begin{eqnarray}
& & V_{ij}(t+1) = [ (1-\epsilon)( f(x_i(t)) - f(x_j(t))) + \nonumber\\
&  & \frac{2 \varepsilon}{N} \sum_{j=N/2+1}^N {g}(x_j(t - \tau))  -  
\frac{ 2 \varepsilon}{N} \sum_{i=1}^{N/2} {g}(x_i(t - \tau)) ]^{2}
\nonumber
\end{eqnarray}
For ideal D state, the synchronization between two nodes which are not directly connected
is independent of the 
delay terms as the coupling terms cancel 
out, and only depends on $\varepsilon$. Hence,
delay does not affect synchronization between the nodes which are not directly 
connected \cite{physicaA2012}, and only comprehends its presence 
for those which are directly connected. As a consequence,
depending upon $\varepsilon$ and $\tau$,
it may either enhance or destroy the synchrony between them.
For instance, in the lower $\varepsilon$ range odd delays lead
to an enhancement of coordination between connected nodes yielding a transition
to SO clusters.
Whereas in middle $\varepsilon$ range, delay destroys 
synchronization between connected nodes yielding D
clusters state.
As indicated by Fig.~\ref{fig_net_clus}, for $\tau=0$, 
the common term in the evolution equation for all
the nodes may be the reason for global synchronization.
Whereas, for $\tau > 0$, the network gets divided into
two parts, one set of
nodes has completely different terms in its evolution equations than
those of the second set.
An important inference of our results
is that in the presence of delay, the dynamical evolution on bipartite network
identifies the underlying network structure
and gives rise to ideal
D clusters for almost all the couplings for $\varepsilon \gtrsim 0.4$.
Note that a previous result on delayed bipartite networks 
concludes that they would lead to worst synchronization \cite{Atay_analytical}, but D clusters
observed here very clearly reveal a very good synchronizing power of the same. 
\begin{figure}
\centerline{\includegraphics[width=0.7\columnwidth]{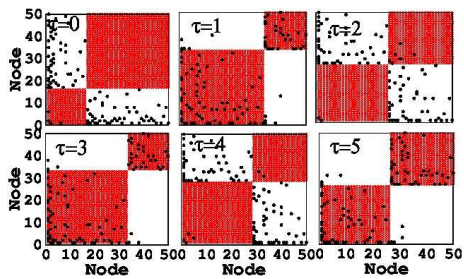}}
\caption{(Color online) Phase synchronized patterns for
coupled circle maps on scale-free networks with $N=50$, $<k> = 2$,
$g(x) = x$ and $\varepsilon = 0.24$.}
\label{Fig_SF_circ}
\end{figure}

In order to demonstrate the robustness of the above
phenomena, we also present results for coupled circle maps.
In Eq.(\ref{cml}), the local
dynamics is defined by the circle map, $f(x) = x + \omega + (p/2\pi)sin(2\pi x)$,
with parameter values taken in a chaotic regime.  
Fig.~\ref{Fig_SF_circ} plots the examples demonstrating the S-D transition, furthermore  different 
$\tau$ values are 
associated with a change cluster pattern as manifested by coupled logistic maps.

We have studied effects of delay on phenomena of phase synchronized 
cluster formation in coupled map networks.
Depending upon $\varepsilon$ values, a change in $\tau$ may lead to a 
change in the phenomenon of cluster
formation, with delays of the same parity being associated with 
the same phenomenon, or favors D clusters for delayed case over undelayed one
which in extreme case of bipartite networks demonstrate robustness of 
D mechanism against change in $\tau$.
Furthermore, different $\tau$ values may lead to an entirely new
pattern of the cluster. For example,
in middle $\varepsilon$ range, different $\tau$ values lead to different
dynamical patterns of dominant D type. Whereas lower $\varepsilon$ values
produce dynamical patterns of dominant D or dominant SO type.

While an enhancement or suppression of complete synchronization as an introduction of delay
was already well investigated in coupled maps models, 
mechanisms of delayed unit to unit interaction were unknown. 
Delay may enhance the coordination among the connected nodes leading to an 
enhancement of synchronization identifying underlying connection topology, 
which had been the main theme of a few recent studies, 
but
observation of a D mechanism behind the cluster formation
in delayed coupled networks is a new insight suggesting 
that delay-induced synchronization
may lead to a completely different relation between functional clusters and topology, than 
relations observed for the
undelayed evolution. 
Our study draws its significance in understanding  synchronization 
in real world networks such as neural networks, where  
clusters are formed due to delayed interactions between neurons 
\cite{brain_web} and may be of D type \cite{sub-population}.
An analysis presented for bipartite and periodic cases
help in discerning a possible impact of $\tau$ on the coupled 
evolution in such systems.  Moreover, a change in patterns of neural activities has 
been found to be
related with brain disorders such as Alzheimer \cite{AD}. Research in the dimension of 
delay-induced patterns might propagate a finer apprehension of the origin and treatment of these diseases.
At fundamental level, a study of phase shift 
synchronization \cite{book_Kurths}, based on phase synchronization
measure considered here, is an aspect to explore in future \cite{note2}.

SJ thanks DST for financial support.


\begin{thebibliography}{99}
\bibitem{network_algorithm}1-D lattices used in the simulation have circular boundary conditions
with each node having $k$ nearest neighbors. Small-world, scale-free and random networks
are obtained by using SW, BA and ER model respectively
\cite{rev-network}. The complete bipartite networks consist of two sets where each node 
of one set is connected with all the nodes of other set. Results are presented for
both the sets having equal number of nodes.
\bibitem{note_inter_intra}In
$N_{inter}$, the coupling between two isolated nodes is not included.
\bibitem{note2}Phase shift synchronization in maps can be defined as follows: maxima (minima) of a 
pair of nodes following a constant time difference for $t > t_0$.
\end{thebibliography}
\end{document}